\documentclass[final,5p,times,twocolumn]{elsarticle} 

\usepackage{hyperref}
\usepackage{lineno}
\usepackage{multirow}
\usepackage{multicol}
\usepackage{gensymb}
\usepackage{amssymb}
\usepackage{tablefootnote}
\usepackage{caption}
\usepackage{subcaption}
\usepackage{wrapfig}
\usepackage{amsmath}
\usepackage[dvipsnames]{xcolor}

\newcommand{\np}{~$\mathrm{n^+}$}

\journal{NIM Section A}
 
\begin{document}
\begin{frontmatter}

\title{DC-coupled resistive silicon detectors for 4D tracking}

\author[address1,address3]{L. Menzio\corref{mycorrespondingauthor}}
\cortext[mycorrespondingauthor]{L. Menzio}
\ead{luca.menzio@edu.unito.it}
\author[address4]{R. Arcidiacono}
\author[address2]{G. Borghi}
\author[address2]{M. Boscardin}
\author[address3]{N. Cartiglia}
\author[address2]{M. Centis Vignali}
\author[address1]{M. Costa}
\author[address5]{G-F. Dalla Betta}
\author[address4]{M. Ferrero}
\author[address2]{F. Ficorella}
\author[address1]{G. Gioachin}
\author[address3]{M. Mandurrino}
\author[address5]{L. Pancheri}
\author[address2]{G. Paternoster}
\author[address3,address1]{F. Siviero}
\author[address3]{V. Sola}
\author[address1]{M. Tornago}

\address[address1]{Università degli Studi di Torino, Torino, Italy}
\address[address3]{INFN, Torino, Italy}
\address[address4]{Università del Piemonte Orientale, Novara, Italy}
\address[address2]{Fondazione Bruno Kessler, Trento, Italy}
\address[address5]{Università degli Studi di Trento, Trento, Italy}

\begin{abstract}

In this work, we introduce a new design concept: the DC-coupled Resistive Silicon Detectors, based on the LGAD technology. This new design intends to address a few known drawbacks of the first generation of AC-coupled Resistive Silicon Detectors (RSD). The sensor behaviour is simulated using a fast hybrid approach based on a combination of two packages, Weightfield2 and LTSpice. The simulation demonstrates that the key features of the RSD design are maintained, yielding excellent space and time resolutions: a few tens of ps and a few microns. In this report, we will outline the optimization methodology and the results of the simulation. We will also present detailed studies on the effects induced by the choice of key design parameters on the space and time resolutions provided by this sensor.

\end{abstract}


\end{frontmatter}



\section{Introduction}

Silicon sensors based on the $n$-in-$p$ LGAD (Low-Gain Avalanche Diode) technology have shown excellent timing capabilities \cite{LGADs}. The AC-coupled Resistive Silicon Detectors (RSD, also called AC-LGAD) \cite{rsd_0} \cite{rsd_1} \cite{rsd_2} couples the LGADs optimal time resolution with outstanding space resolution, making AC-LGAD  one of the emerging technologies for 4D tracking \cite{Arci1}. However, a few important issues have emerged that can impact their performance: (i) signals spread past the nearest pads, (ii) the number of pads seeing a signal depends on the hit position, (iii) the signal is bipolar, and (iv) the signal baseline in large and/or irradiated sensors might fluctuate. In this paper, we introduce the DC-coupled RSD (DC-RSD), a new approach to resistive read-out sensors that intends to address these drawbacks by implanting the read-out electrodes directly  into the $n^+$ resistive layer. This design yields to controlled charge sharing patterns, unipolar signals, and absence of baseline fluctuations.

\section{Simulations approach}
\begin{figure}[!htb]
    \centering
    \includegraphics[width=0.35\textwidth]{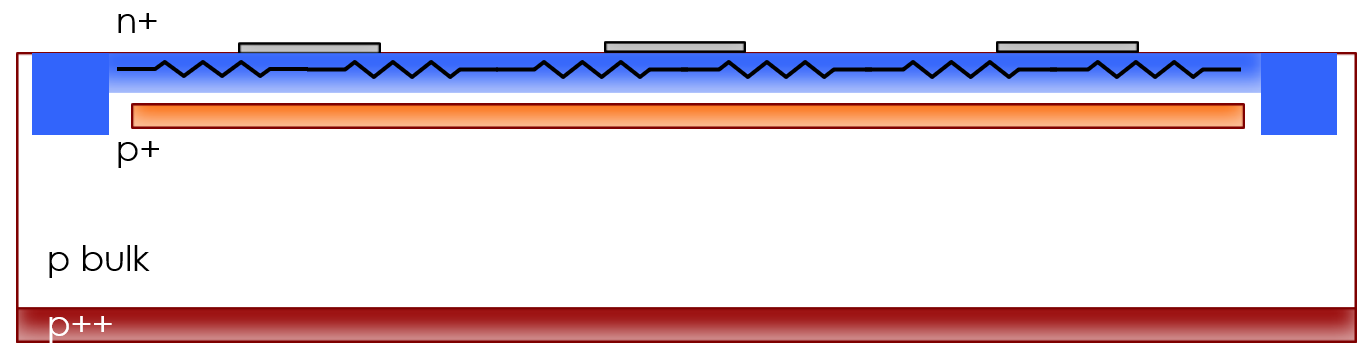}
    \captionsetup{belowskip=-15pt}
    \caption{Schematics of a DC-RSD.}
    \label{fig:DCRSD-scheme}
\end{figure}

A basic DC-RSD consists of a LGAD design with a single unsegmented p+ gain layer and a n+ layer on which the output signal is read out through metallic pads (grey in Fig. \ref{fig:DCRSD-scheme}).

Traditionally, silicon sensors simulations are performed exploiting TCAD-based software. This approach has the benefit of being very accurate, although it is very demanding in terms of computing time.
The time needed for a 3D simulation increases very rapidly with the simulated volume, for example,  the simulation of a 100 x 100 $\mu$m$^2$ pixel with a 50 $\mu$m thick active area takes about 10 hours. The simulation of RSD sensors is particularly challenging since the signal spreads over large areas, requiring a 3D simulation of areas up to a few mm$^2$.  In order to perform a large number of simulations to explore the best DC-RSD design, we opted for a different type of approach, splitting the simulation process into two steps: 

\begin{enumerate}
    \item the passage of a particle through the sensor bulk and the consequent signal formation
    \item the charge spreading over  the \np layer towards the read-out contacts placed at virtual ground.  
\end{enumerate}
In the first step, we employed the Weightfield2 (WF2) \cite{weightfield2} software package, while in the second step  the LTspice schematics simulator \cite{ltspice}.

\subsection{Weightfield2}
\label{par:weighfield}
WF2 has been tuned to reproduce accurately the signal generated by a
 Minimum Ionizing Particle (MIP) impinging on a silicon sensor. Since the initial signal formation in DC-RSD is similar to that of a standard LGAD sensor, the current stimulus can be obtained simulating the signal in a standard LGAD of the same thickness (50 \( \mu \)m). For this analysis, the LGAD parameters used in WF2 were those of the  W13 from the FBK UFSD3.2 production \cite{UFSD3.2}, biased with a voltage of 200 V. 
The current signal generated in this way  is recorded and employed in the next step. 

\subsection{LTspice}

\begin{wrapfigure}{l}{0.25\textwidth}
    \includegraphics[width=0.25\textwidth]{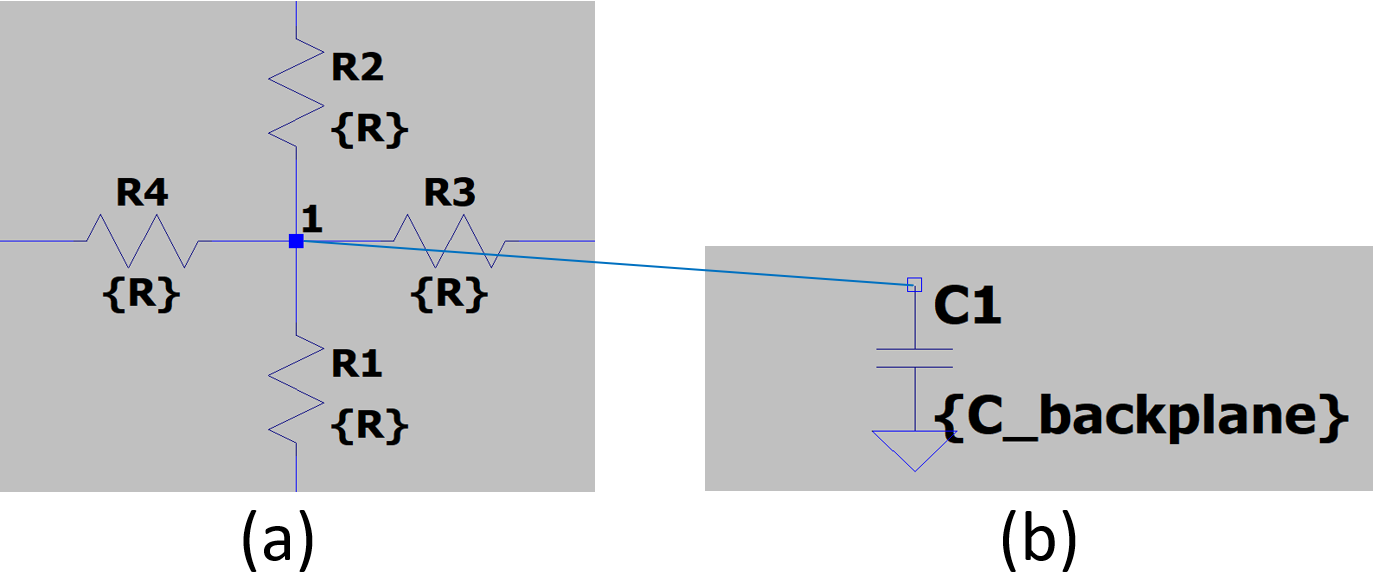}
    \caption{Scheme of the LTspice simulation fundamental block.}
    \label{fig:fundamental_block_ltspice}
\end{wrapfigure}

The signal spread over the sensor surface is simulated using the LTspice schematics simulation tool by Analog Devices. 
In this approach, the key elements of DC-RSD are represented by electrical components. The resistive plane is modeled by a network of resistors, the sensor bulk by capacitors,  and the front-end electronics by resistors that approximate the read-out input impedance.  The fundamental block of the detector modelization (see Fig.\ref{fig:fundamental_block_ltspice}) is a node connected to four resistors (a) and a capacitor connected to ground (b). 


Typical values are in the order of a few k$\Omega$/sq for the sheet resistivity and a few fF for the bulk capacitance. The value of the capacitance has been computed using as model a parallel plate capacitor, i.e. $C =\epsilon_0 \epsilon_r \frac{A}{d}$ where $A$ is the hypotetical area  covered by the node, $d$ the thickness of the detector,  $\epsilon_0 \epsilon_r$ the silicon dielectric permittivity. Unless otherwise specified, we simulated a 340 $\mu$m-wide sensor with 15x15 blocks.

\section{Position Reconstruction}
The performance of DC-RSDs is evaluated by measuring their spatial precision.  This is obtained by injecting  the signal simulated with WF2 (see paragraph \ref{par:weighfield}) in each node of the resistive plane and comparing the reconstructed positions to the real positions.

The position is reconstructed by measuring the amplitude imbalance of the currents measured at the four corners of the resistive plane (the 4 red dots in the picture on the left of Fig. \ref{fig:first_reconstruction}). The x-y coordinates are determined as:

\begin{equation}
    \label{eq:amplitude_imbalance}
    \begin{aligned}
    x= \frac{A_2 + A_3 - A_1 - A_4}{A_{tot}} \\
    y= \frac{A_1 + A_2 - A_3 - A_4}{A_{tot}},
    \end{aligned}
\end{equation}
where $A_i$ is the signal  amplitude at the $i$ corner  and $A_{tot}$ the sum of the four amplitudes. Such position reconstruction approach has been maintained constant throughout all simulations. 

In order to match the front-end simulation  to the real-life situation, the signal at each read-out pad is amplified with a trans-impedance of 4700 $\Omega$, typical of lab amplification board. A gaussian smearing with 2 mV RMS is added to each read-out signal to reproduce the effect of the noise.

\section{Optimization of the DC-RSD Design}
\label{par:goodness}
In this section, we present 3 different designs of DC-RSD: (i) read-out pads at the 4 corners, (ii) read-out pads at the 4 corners connected via a grid of constant resistors, and (iii) read-out pads at the 4 corners connected via a grid of variable resistors.  

In order to evaluate the performances of the configurations, two goodness-of-reconstruction parameters have been defined:
\begin{itemize}
    \item \textbf{Average point accuracy $d$}, defined as in the following equation (eq. \ref{eq:mean_dist})
    \begin{equation}
        \label{eq:mean_dist}
        d_i = \frac{\sum\limits_{n = 0}^N |\Vec{x_r^{n,i}}-\Vec{x_o^i}|}{N}, \;\; d = \frac{\sum\limits_{i=0}^{N_{nodes}} d_i}{N_{nodes}}
    \end{equation}
    where $i$ is the node index, $\Vec{x_r^{n,i}}$ the reconstructed position and $\Vec{x_o^i}$ the injection one, $N$ the number of trials per node, $N_{nodes}$ the total number of nodes. 
    \item \textbf{Mean reconstructed position dispersion $\sigma$}
    \begin{equation}
        \label{eq:sigma}
        \Vec{x_{avg}^i} = \frac{\sum\limits_{n = 0}^N\Vec{x_r^{n,i}} }{N}, \;\; \sigma = \frac{1}{N_{nodes}} \sum\limits_{i=0}^{N_{nodes}} \frac{\sum\limits_{n=0}^N |\Vec{x_r^{n,i}}-\Vec{x_{avg}^i}| }{N} 
    \end{equation}
\end{itemize}
The best configurations are those that minimize both parameters.

In addition to the spatial resolution, we also want to minimize the time resolution. This is obtained by ensuring that the signals have high $dV/dt$, i.e. have high amplitude and short rising time, for equal noise level \cite{LGADs}. 

\subsection{Read-out pads at the 4 corners}

\begin{figure}[h]
    \centering
    \includegraphics[width=0.48\textwidth]{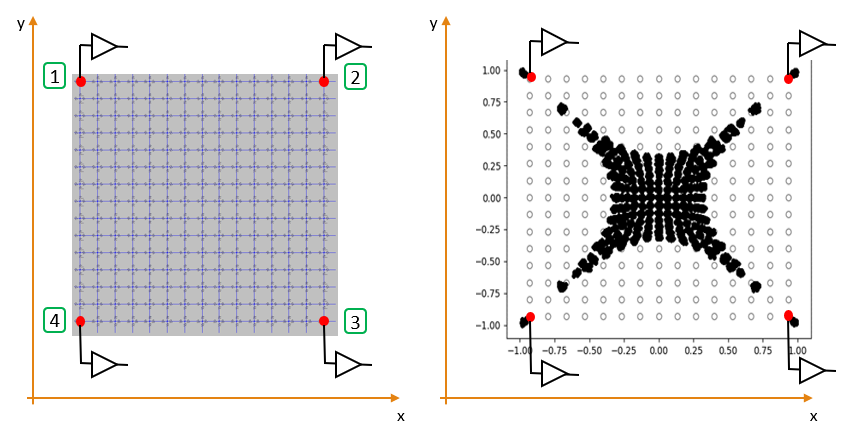}
    \caption{Left:  simulated configuration. Right: true (empty circles) and reconstructed positions (full circles) of this simulation.}
    \label{fig:first_reconstruction}
\end{figure}

This first geometry has been implemented with a sheet resistivity typical of the RSD1 production, i.e. about 1 k$\Omega$/sq, and the capacitance of each node is 1.29 fF,  tuned to represent a 340 $\mu$m-wide sensor. The left pane of Fig. \ref{fig:first_reconstruction} shows the simulated configuration while  on the right pane  the true (empty circles) and  reconstructed (full  circles) positions of the signals. The distortion visible on the right picture in Fig. \ref{fig:first_reconstruction} is typical of resistive detectors and it has already been described in literature \cite{Microcat}.

In order to maintain the number of pads seeing a signal as constant as possible, it is crucial to insulate the square delimited by the four read-out pads. In this way it is more likely to have high enough signals, i.e. whose amplitude are grater than the noise level and can participate to the position reconstruction.

\subsection{Read-out pads at the 4 corners connected via a grid of constant resistors }
\label{par:low_res}
\begin{figure}[h]
    \centering
    \includegraphics[width=0.38\textwidth]{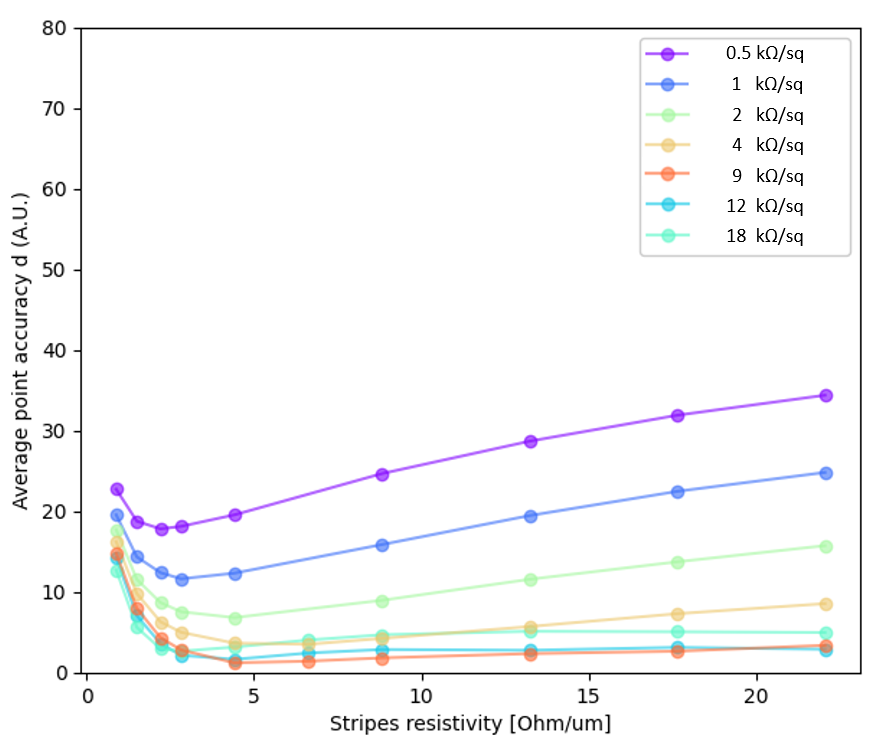}
    \caption{Mean distance as a function of the strips linear resistance and the sheet resistivity (see labels).}
    \label{fig:mean}
\end{figure}

\begin{figure}[h]
    \centering
    \includegraphics[width=0.38\textwidth]{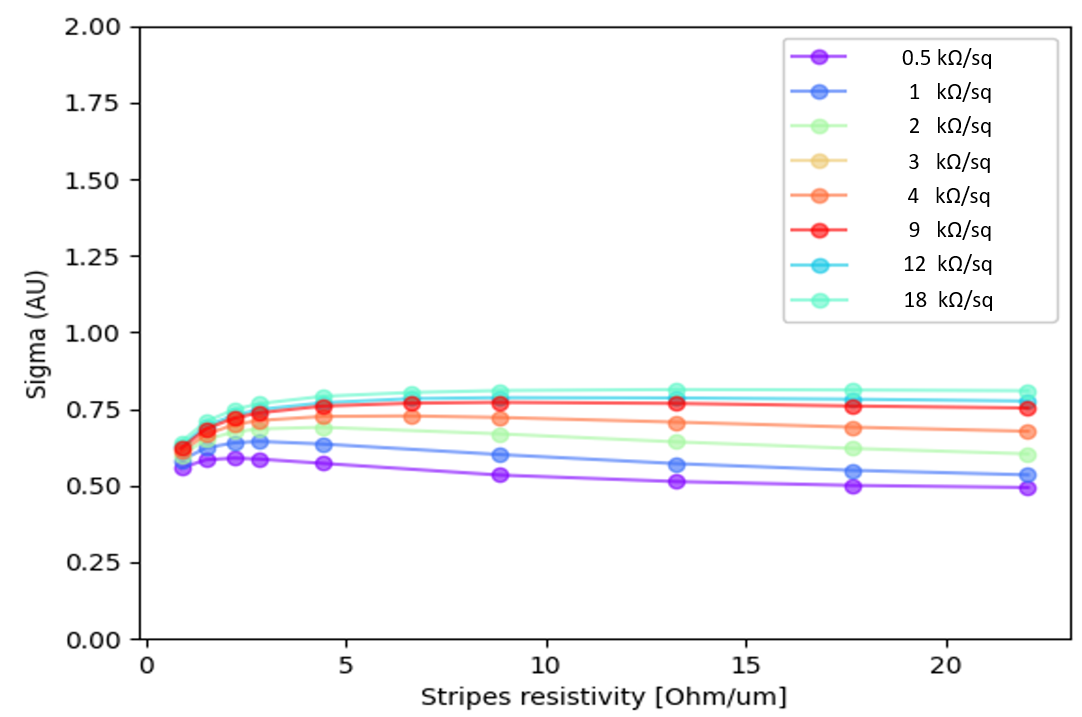}
    \caption{Sigma parameter as a function of the strips linear resistance and the sheet resistivity (see labels).}
    \label{fig:sigma}
\end{figure}

By looking at the reconstruction of Fig. \ref{fig:first_reconstruction}, it is clear that the reconstructed points tend to cluster in the centre of the plane. This effect can be strongly reduced by adding resistive strips connecting the read-out electrodes; such strips have resistivity per micron lower than the sheet resistivity.  The presence of these strips equalizes charge sharing, strongly improving the reconstruction.
\begin{wrapfigure}{l}{0.25\textwidth}
    \includegraphics[width=0.25\textwidth]{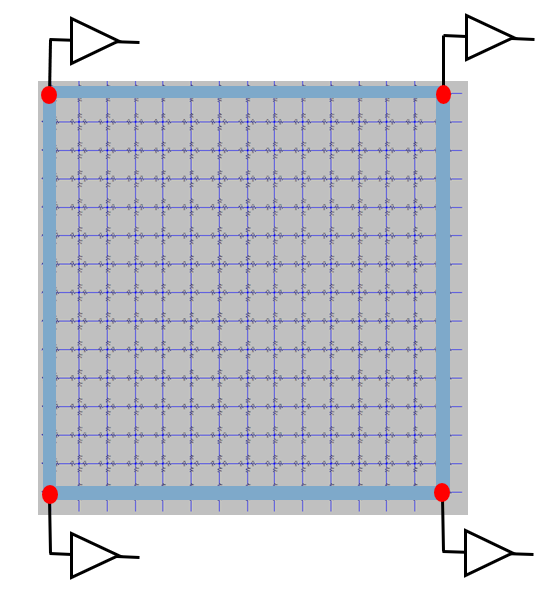}
    \caption{Layout of the low-resistivity strips simulations: low-resistivity strips connect the read-out electrodes.}
    \label{fig:low_res_layout}
\end{wrapfigure}
It must be noted that the resistance between the electrodes should ensure isolation between the read-out amplifiers. Because of this, the total resistance of each strip has been kept above 10 times the electronics input impedance. The low resistivity strips are simulated in the LTspice environment as a single line of resistors with a resistance lower than those of the resistive sheet. Different values of sheet resistivity and strips resistance were tried. The results are reported in Fig. \ref{fig:sigma} and Fig. \ref{fig:mean}. Each point on the graphs is obtained with 225 independent signal injections (simulations), one for each node.    

From the trend of the two parameters one could assess that: (i) the best value of the strip linear resistance increases weakly  with the sheet resistivity. Overall, the minimum $d$  is localized around values of strip linear resistance $R_{strip} = 1-3 \; \Omega /\mu m$; (ii) higher sheet resistivities lead to smaller dispersion; (iii) the mean dispersion $\sigma$ is practically constant in all configurations.

\begin{figure}[h]
    \centering
    \includegraphics[width=0.40\textwidth]{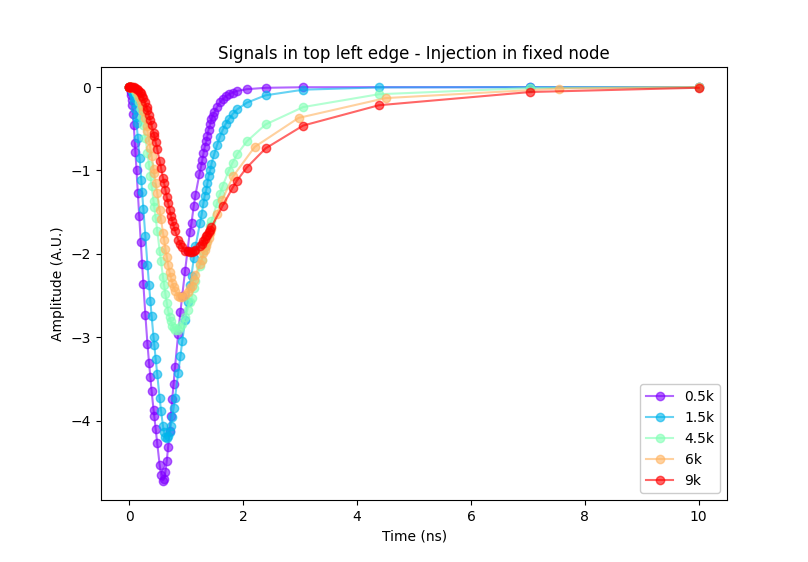}
    \caption{Signals picked up in a single electrode in different sized simulated sensors.}
    \label{fig:signal_comp_ressheet}
\end{figure}

Then, keeping the strip resistance at a value of $R_{strip} = 3 \; \Omega /\mu m$, we studied the shape of the signals as a function of the sheet resistivity. This analysis shows that the shape of the output signals tends to become smaller and longer for high sheet resistivity, worsening the timing capabilities (see Fig. \ref{fig:signal_comp_ressheet}. 

\begin{figure}[h]
    \centering
    \includegraphics[width=0.28\textwidth]{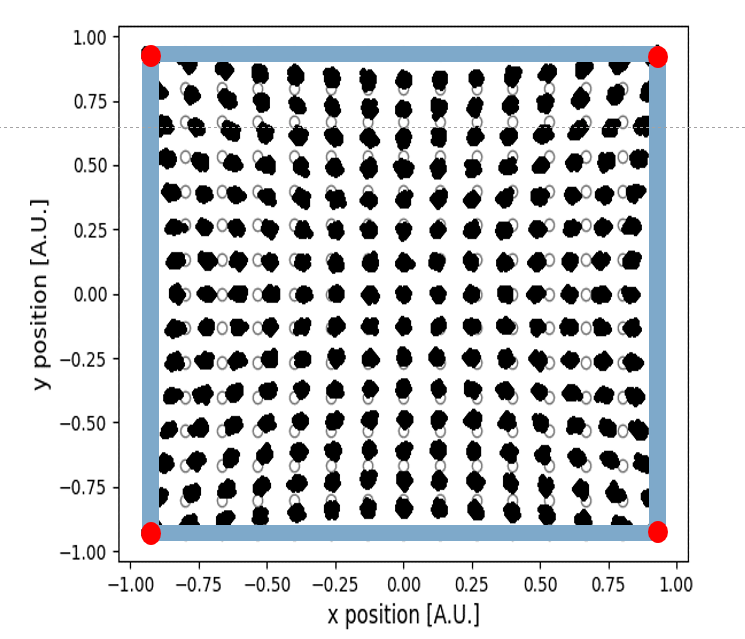}
    \caption{Reconstruction pattern for a the best combination of sheet-strip resistivity (see text for further details).}
    \label{fig:low_res_strip_rec}
\end{figure}

Combining the spatial and temporal analysis, we opted for a sheet resistivity $R_{sheet} = 3 \; k\Omega / sq$ and a strip resistivity of about $R_{strips} = 2.5 \; \Omega / \mu m$ for a 340 $\mu m$-wide detector. Such configuration ensures that the signal is shared between four read-out pads and provides the reconstruction pattern visible in Fig. \ref{fig:low_res_strip_rec}. 

\subsection{Read-out pads at the 4 corners connected via a grid of variable resistors.}
\label{par:var_strips}
\begin{figure}[h]
    \centering
    \includegraphics[width=0.48\textwidth]{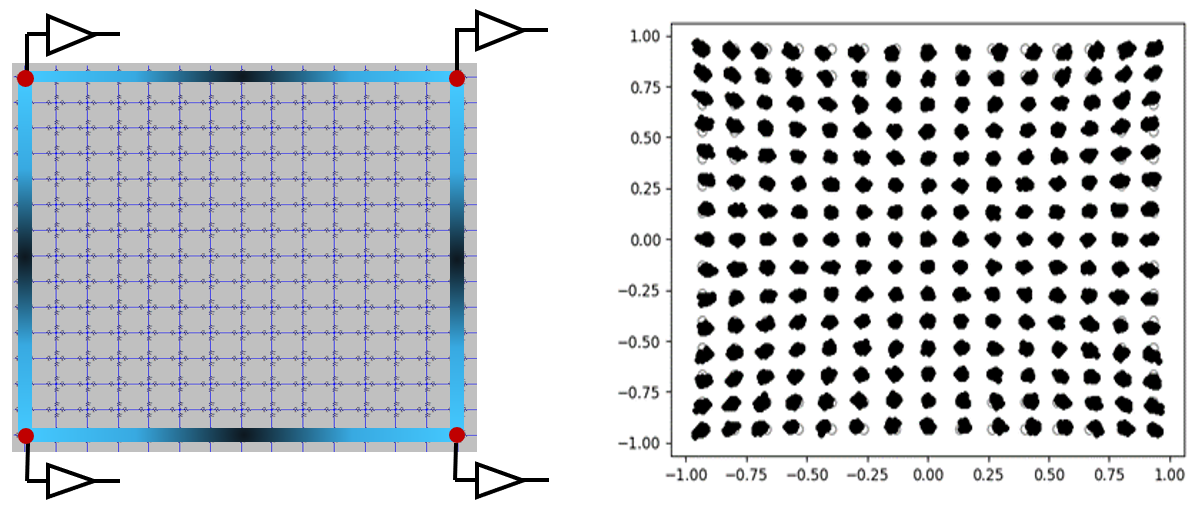}
    \caption{Left: sketch of the variable resistance strips solution. Right: best reconstruction obtained in this configuration (see text for more detailed information).}
    \label{fig:variable_strip}
\end{figure}

Fig. \ref{fig:low_res_strip_rec} shows that the reconstructed positions are still systematically shifted with respect to the true positions, even when using the optimal strip resistance. The points tend to slightly cluster toward the centre of the sensor: this indicates that too much charge is reaching the side further away from the injection point. To compensate for this effect, we introduced a  modulation of the strip resistance: increasing its value in the central part of the strip we decreased the flow of charge to the far-away read-out electrode. A sketch of this solution is shown on the left pane of Fig. \ref{fig:variable_strip}. 
We tried different maximum and minimum values of the strip resistance, obtaining best results with the configuration characterized by a resistive sheet  $R_{sheet} = 3 \; k\Omega / sq$ and strip resistivity in the range $R_{strip} = 1\div10 \;  \Omega / \mu m $ with a variation of 25\% - 50\% along the strip length. It can be noted how this configuration yields reconstructed points that almost overlap with the original ones.

\subsection{The effect of pixel size}
\label{par:pad_sizes}
In the last part of the analysis we studied how the size of the detector impacts the performance. This is obtained by tuning the backplane capacitance while maintaining constant the other parameters. The explored dimensions range from 130 $\mu m$ to 1 $mm$. The signals maintain a sharp rising edge and almost the same amplitude. This study agrees well with our experimental finding that signal propagation on a resistive sheet with $R_{sheet} = 1-3 \; k\Omega / sq$ does not significantly changes the signal shape \cite{N_Trento, L_Trento}.

Therefore, we foresee the possibility of having a DC-coupled sensor with large pixels (up to 1 x 1 mm$^2$) yielding the spatial resolution typical of RSD2 (better than 5\% of the pixel size) and the time resolution (30-40 ps) typical of LGAD-based sensors. 

\subsection{TCAD Simulations and Future Production}
In parallel to this analysis, a study using a full TCAD simulation is in progress in order to launch a DC-RSD production at the Fondazione Bruno Kessler in Summer 2022.
\section{Conclusions}
\label{par:conclusions}
In this paper, we propose an evolution  in the design of silicon sensors with resistive read-out: the DC-coupled Resistive Silicon Detector (DC-RSD). In DC-RSD,  the read-out electrodes are directly embedded  in the \np resistive plane and the presence of low-resistivity strips connecting the electrodes improves spatial accuracy. In order to maximize the sensor performances,  we introduced the concept of variable strip resistance, further improving the position reconstruction. From these sensors, we expect an excellent time (30-40 ps) and space (20-30 $\mu m$) resolution even with sensors with  very large pitch ( up to~ 1 mm). The first DC-RSD production by Fondazione Bruno Kessler (FBK) is planned for Summer 2022.

\section{Acknowledgements}
We kindly acknowledge the following funding agencies and collaborations: INFN – FBK agreement on sensor production; Dipartimenti di Eccellenza, Univ. of Torino (ex L. 232/2016, art. 1, cc. 314, 337); Ministero della Ricerca, Italia, PRIN 2017, Grant 2017L2XKTJ – 4DinSiDe; Ministero della Ricerca, Italia, FARE,    Grant R165xr8frt\_fare



\end{document}